\documentclass[10pt,a4paper,twoside]{article}

\usepackage{epsfig}
\usepackage{baltlat6}
\usepackage{array}
\usepackage{here}
\begin{document}
\ \
\vspace{-0.5mm}
\setcounter{page}{1}
\vspace{-20mm}

\titlehead{Baltic Astronomy, vol.\,00, 00}

\titleb{Hydrodynamic Studies of the Evolution of Recurrent, Symbiotic, and Dwarf Novae:
The White Dwarf Components are Growing in Mass}

\begin{authorl}
\authorb{S. Starrfield}{1},
\authorb{F. X. Timmes}{1},
\authorb{C. Iliadis}{2},
\authorb{W. R. Hix}{3},
\authorb{W. D. Arnett}{4},
\authorb{C. Meakin}{5}, \&
\authorb{W. M. Sparks}{5}
\end{authorl}

\begin{addressl}

\addressb{1}{School of Earth and Space Exploration, Arizona State University, Tempe, AZ 85287-1404;
starrfield@asu.edu; fxt44@mac.com}
\addressb{2}{Dept. of Physics \& Astronomy,  University of North Carolina, Chapel Hill, NC 27599-3255; iliadis@unc.edu}
\addressb{3}{Dept. of Physics and Astronomy, University of Tennessee, Knoxville, TN 37996-1200;
raph@utk.edu}
\addressb{4}{Dept. of Astronomy, University of Arizona, Tucson, AZ, 85721; \\darnett@as.arizona.edu}
\addressb{5}{Los Alamos National Laboratory, Los Alamos, NM, 87545; \\ casey.meakin@gmail.com;
warrensparks@comcast.net }
\end{addressl}

\submitb{Received: 2011 September 1; accepted: }

\begin{summary}
Symbiotic binaries are systems with white dwarfs (WDs) and red giant companions.  
Symbiotic novae are those systems in which thermonuclear eruptions occur on the WD components.   
These are to be distinguished from events driven by accretion disk instabilities analogous to Dwarf 
Novae eruptions in Cataclysmic Variable outbursts.   Another class of Symbiotic systems are
those in which the WD is extremely luminous and it seems likely that quiescent nuclear burning is ongoing on the accreting WD.   
A fundamental question is the secular evolution of the WD.  Do the repeated outbursts or
quiescent burning in these accreting systems cause the WD to
gain or lose mass? If it is gaining mass, can it eventually reach the Chandrasekhar Limit and
become a supernova (a SN Ia if it can hide the hydrogen and helium in the system)?  
In order to better understand these systems, 
we have begun a new study of the evolution of Thermonuclear Runaways (TNRs) in the accreted envelopes of WDs  
using a variety of initial WD masses, luminosities, and mass accretion
rates.  We use our 1-D hydro code, NOVA, which
includes the new convective algorithm of Arnett, Meakin, and Young, the Hix and Thielemann nuclear reaction solver, the lliadis reaction rate library, the
Timmes equation of state, and the OPAL opacities.
We assume a solar composition (Lodders abundance distribution)
and do {\em not} allow any mixing of accreted material with core material. This assumption strongly influences our results.

We report here (1) that the WD grows in mass for all simulations so that ``steady burning'' does not occur, and
(2) that only a small fraction of the accreted matter is
ejected in some (but not all) simulations.    We also find that the accreting systems, before thermonuclear runaway, are
too cool to be seen in X-ray searches for SN Ia progenitors. 
\end{summary}

\begin{keywords}
stars: white dwarfs, close binaries, dwarf novae, interiors, novae, cataclysmic variables, supernovae
\end{keywords}

\resthead{Hydrodynamic Studies of Symbiotic Novae}
{Starrfield, Timmes, Iliadis, Hix, Arnett, Meakin, Sparks}

\sectionb{1}{INTRODUCTION}

One of the fundamental questions concerning the long term evolution of Symbiotic Binaries is whether
or not the mass of the white dwarf (WD) is growing and, if growing, at what rate per year?   The answer to this question 
determines if the mass of the WD can reach the Chandrasekhar Limit and then explode as a Supernova Ia (SN Ia).
Because SN Ia can become extremely bright at maximum light and, thus, be detected at cosmologically
useful distances, they have become crucial to our understanding 
of the evolution of the Universe.  In addition, because they are thought to occur as the 
consequence of degenerate carbon burning in the core of a carbon-oxygen WD, which has as its
main nucleosynthetic product $^{56}$Ni, they are also important to the Galactic chemical evolution of the iron group elements.
However,  there is no general agreement on the progenitors of the explosion.
Reviews of the various proposals for the search for SN Ia progenitors (Branch, et al.
1994),  and producing a SN Ia and their implications can be found in Hillebrandt \& Niemeyer (2000),
Leibendgut (2000, 2001), and Nomoto et al. (2000).

\begin{figure}[]
\vspace{-10mm}
\vbox{
\centerline{\psfig{figure=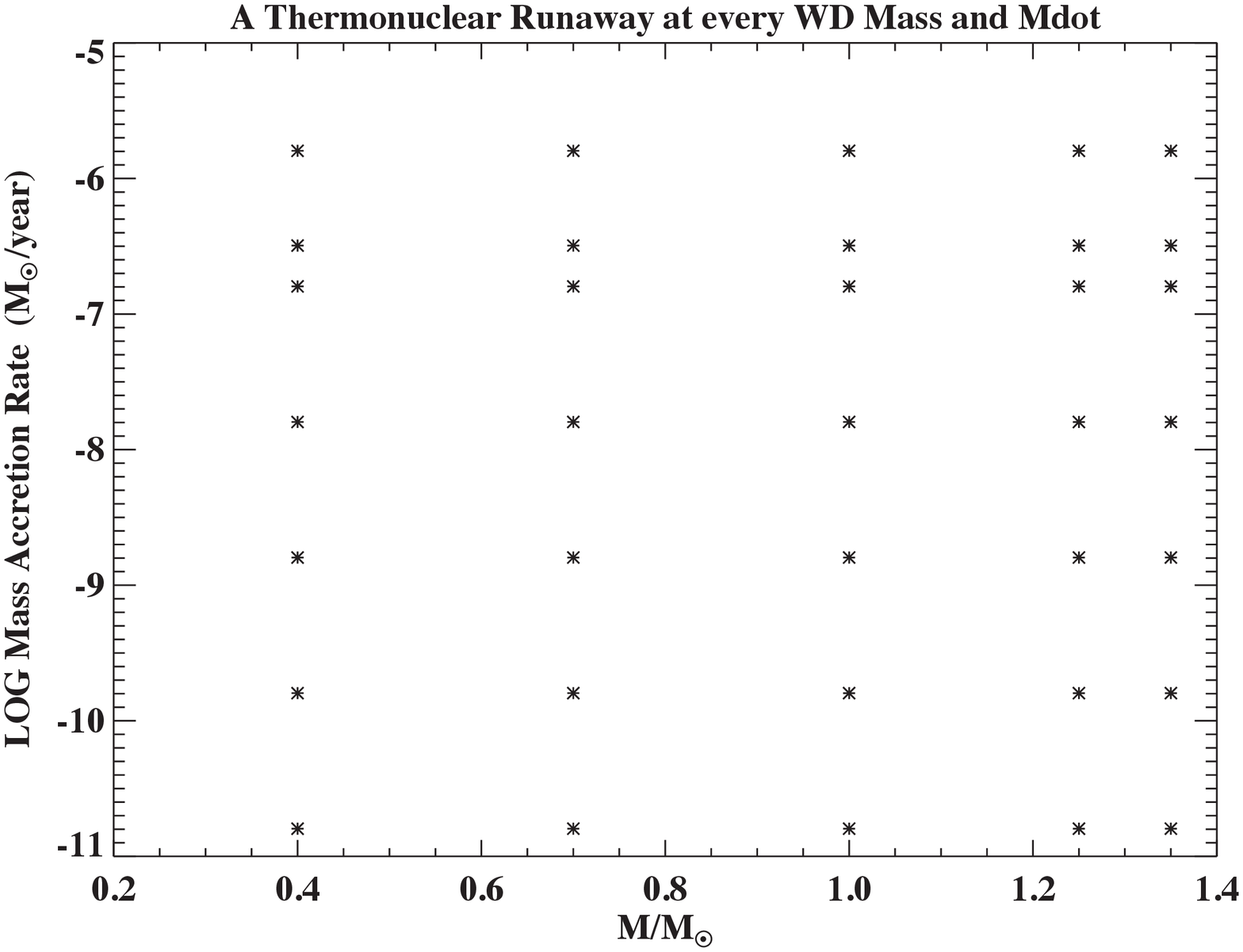,width=100mm}}
\vspace{-5mm}
\captionb{1}
{WD mass and Log \.M for each of the evolutionary sequences that we calculated.  
Since each point represents the two initial luminosities we used, there are
70 sequences shown here.  Each of these
exhibited a TNR.  In no case did ``steady burning'' occur.}}
\end{figure}

The two major suggestions for the SN Ia explosion are the single degenerate and double degenerate scenarios.  In 
the standard paradigm single degenerate scenario (SD), it is proposed that 
a WD in a close binary system accretes material from its companion and grows to the 
Chandrasekhar Limit.  As it nears the Limit, an explosion is initiated 
in the core.   In contrast, the double degenerate scenario (DD) 
requires the merger or collision of two WDs to produce the observed explosion.   While for many years the 
SD scenario was the more prominent, a number of concerns have now led to major
efforts to better understand the DD scenario, in spite of the fact that the SD scenario is
capable of explaining most of the observed properties of the SN Ia explosions via the delayed detonation
model  (Khokhlov 1991; Kasen et al. 2009; Kromer et al. 2010; Woosley \& Kasen 2011, and references therein).

In this paper we explore the SD scenario for SN Ia progenitors which is based on the suggestion of Whelan and Iben (1973)
that the outburst occurs in a close binary system that contains a WD and another star.  
Since the WD is accreting material from a secondary, virtually every type of close binary has been suggested as a SN Ia progenitor but
one of the defining characteristics of a SN Ia explosion is the absence of hydrogen
or helium in the spectrum at any time during the outburst or decline.  This absence appears to
rule out most of the proposed close binary progenitors.   While a recent review by Tornamb\'e \& Piersanti (2005) is quite negative
about accretion onto a WD being the cause of the SN Ia explosion, 
Mazzali et al. (2007) are not as negative.  A recent and detailed review of SN Ia's can be found in Howell (2011). 

In support of the SD scenario, observations of V445 Pup (Nova 2000) imply that it was a helium nova (helium accretion onto a WD) because there
were no signs of hydrogen in the spectrum at any time during the outburst, but there were 
strong lines of carbon, helium, and other elements (Woudt et al. 2009 and references therein).  The secondary is thought to be a hydrogen
deficient carbon star (Woudt et al.\  2009).  Its existence, therefore, implies that there are
binaries in which hydrogen is absent. 
In addition, there are observational searches for faint SN Ia and other objects that might fill the ``gap'' in brightness
between SN Ia and bright Classical Novae  (Kasliwal et al.\  2011) which provide further motivation for our work.

\begin{figure}[]
\vspace{-5mm}
\vbox{
\centerline{\psfig{figure=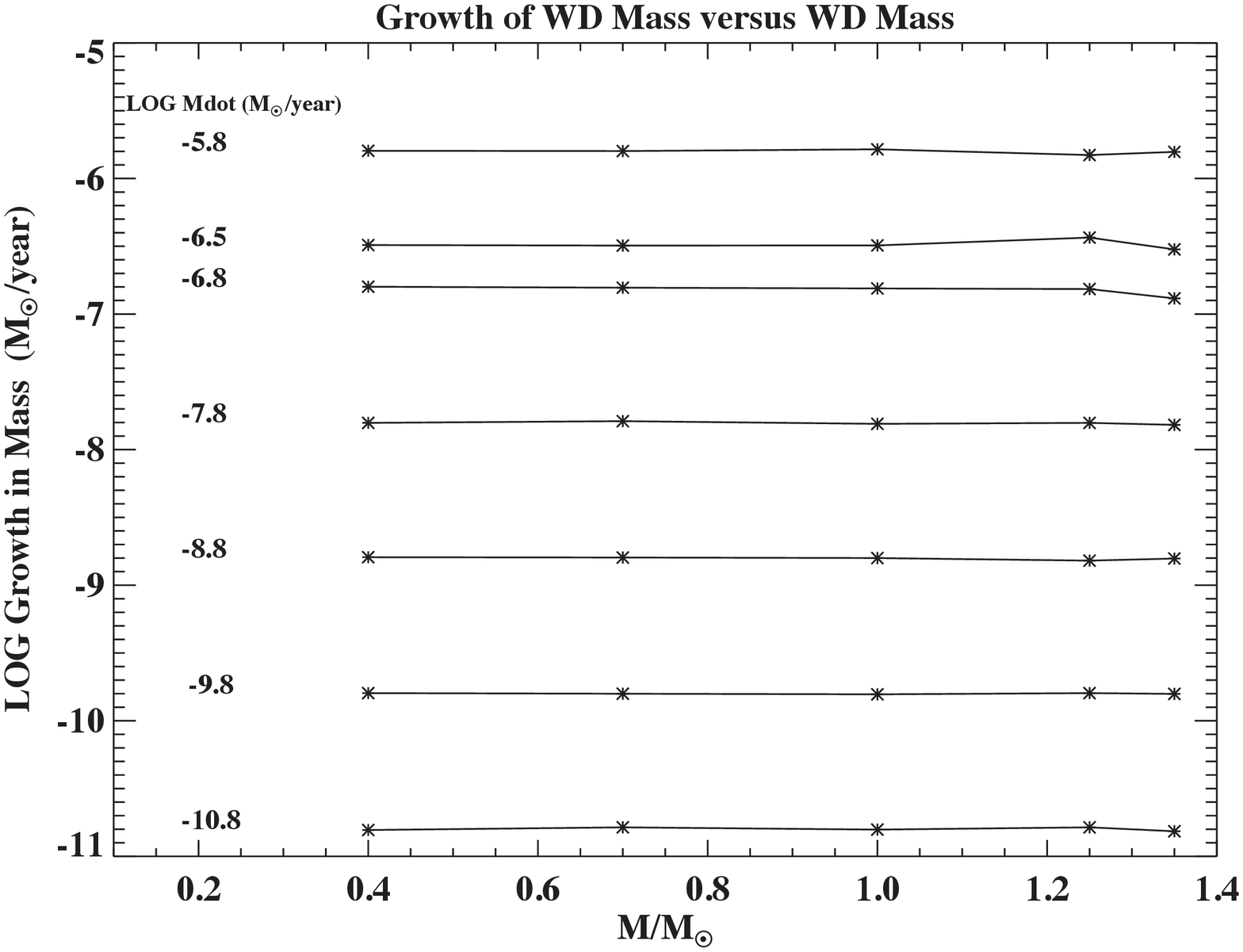,width=100mm}}
\vspace{-5mm}
\captionb{2}
{ The log of the difference between the mass accreted and the mass lost.  
We display the log of the growth in mass  (in units of M$_\odot$yr$^{-1}$)
 as a function of WD mass for each of our simulations.  Each point is the amount of accreted (less ejected) mass divided by the time to 
 reach the TNR for the given simulation.  The lines connect the points for the same \.M and we give the log of  \.M along a column on
 the left of the figure. }}
\end{figure}

\sectionb{2}{THE NOVA CODE}

We use our one-dimensional (1-D) hydrodynamic computer code (NOVA) to study the accretion of {\it solar composition} material
(Lodders 2003) onto WD masses of 0.4M$_\odot$, 0.7M$_\odot$, 1.0M$_\odot$, 1.25M$_\odot$, and 1.35M$_\odot$.  
We use two initial WD luminosities  ($4 \times 10^{-3}$ L$_\odot$ and $10^{-2}$L$_\odot$) and seven mass accretion rates ranging 
from $2 \times 10^{-11}$M$_\odot$ yr$^{-1}$ to $2 \times 10^{-6}$M$_\odot$ yr$^{-1}$.  
Our initial conditions are chosen to mimic those observed for the broad variety of CVs
just as we have done in our previous studies of the Classical and Recurrent Nova outbursts.  We  use the updated 
version of NOVA (Starrfield et al. 2009, and references therein) that includes a nuclear
reaction network that has since been extended to 187 nuclei (up to $^{64}$Ge).  
We use the nuclear reaction rate library of Iliadis et al. (see Starrfield, et al. 2009). 
NOVA now includes the latest microphysics (equations of state, opacities, and electron conduction) and the Arnett, Meakin, \& Young (2010) algorithm for mixing-length convection (see also Meakin and Arnett 2007).  
We find that using the new microphysics produces quantitative but not qualitative changes in our Classical Nova simulations.
The simulations we report here were done with 150 Lagrangian mass zones with the surface zone mass less than $\sim 10^{-9}$M$_\odot$.  Previously,
we used 95 mass zones with a surface mass about 100 times larger.  
The mass of the surface zone determines the maximum time step that can be taken and 
some of our simulations required 4 million time steps.

\begin{figure}[]
\vspace{-5mm}
\vbox{
\centerline{\psfig{figure=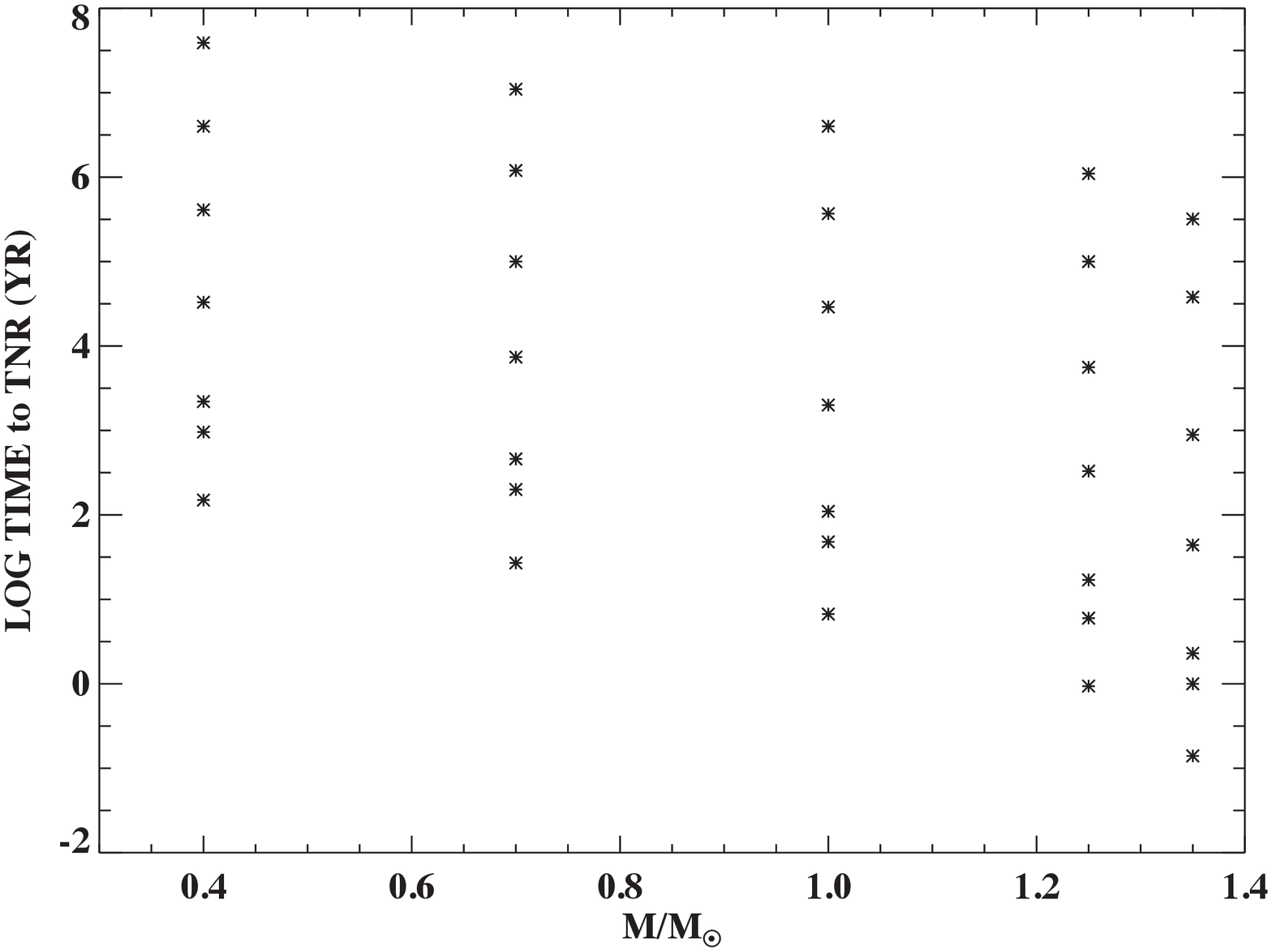,width=100mm}}
\vspace{-5mm}
\captionb{3} {The log of the accretion time to the TNR as a function of WD mass.  Each of the data
points is for a different \.M and the value of \.M increases downward for each WD mass.  The accretion time,
 for a given \.M decreases with WD mass because it takes less mass to initiate the TNR as the WD mass
 increases.}}
\vspace{-2mm}
\end{figure}

\sectionb{3}{MOTIVATION}

In addition to studying the consequences of accretion at a variety of rates onto a variety of WD masses, we are
interested in comparing our simulations to earlier results shown in the \.M - M$_{\tt WD}$ plane (Fujimoto 1982a,b).
A current version of this plot is given as Figure 5 of Kahabka \& van den Heuvel (1997) so we do not reproduce it here.   
This plot has 3 regions on it.  For the lowest mass accretion rates, at all WD masses,  it is predicted that 
accretion results in flashes which are normally expected to resemble those of Classical Novae (Starrfield, Iliadis, \& Hix 2008).  For the
highest mass accretion rates, it is predicted that the radius of the WD rapidly expands to red giant dimensions and accretion is halted. 
There is a third regime, intermediate between these two, where the material is predicted to steadily burn at the accretion rate
(Fujimoto 1982b; see his Figure 9 but notice that it is upside down from the way it is now plotted).  
This \.M is nominally$\sim 3 \times 10^{-7}$M$_\odot$ yr$^{-1}$ and has a small increase with increasing WD mass.   The implications of this
plot is that most of the observed systems are accreting at rates that are not within the ``steady burning'' regime and, therefore, the
WDs cannot be growing in mass.  However, those systems that are accreting at the ``steady burning'' rate are evolving horizontally
towards higher WD mass and, by some unknown mechanism, are stuck in this mass accretion range.  The Super Soft Sources in the LMC are
predicted to be in the ``steady burning'' regime (van den Heuvel et al. 1992).  This scenario implies that since most observed systems
are not accreting at this value the WDs cannot be SN Ia progenitors.  
While, it is important to test this picture, we emphasize that this plot as given
implies that the only parameters that affect the evolution of a WD are its mass and \.M.  It does not take into account 
the chemical composition of the accreting material, the chemical composition of the underlying WD, or if/when mixing of 
accreted material with core material has taken place.  
In addition, it does not take into account the thermal structure of the underlying WD 
and the effects of previous (or continuing) outbursts on the thermal and compositional structure of the WD.  
It is well known that all these parameters affect the evolution of the WD (Starrfield 1989; Starrfield, Iliadis, \& Hix  2008).

\begin{figure}[]
\vspace{-5mm}
\vbox{
\centerline{\psfig{figure=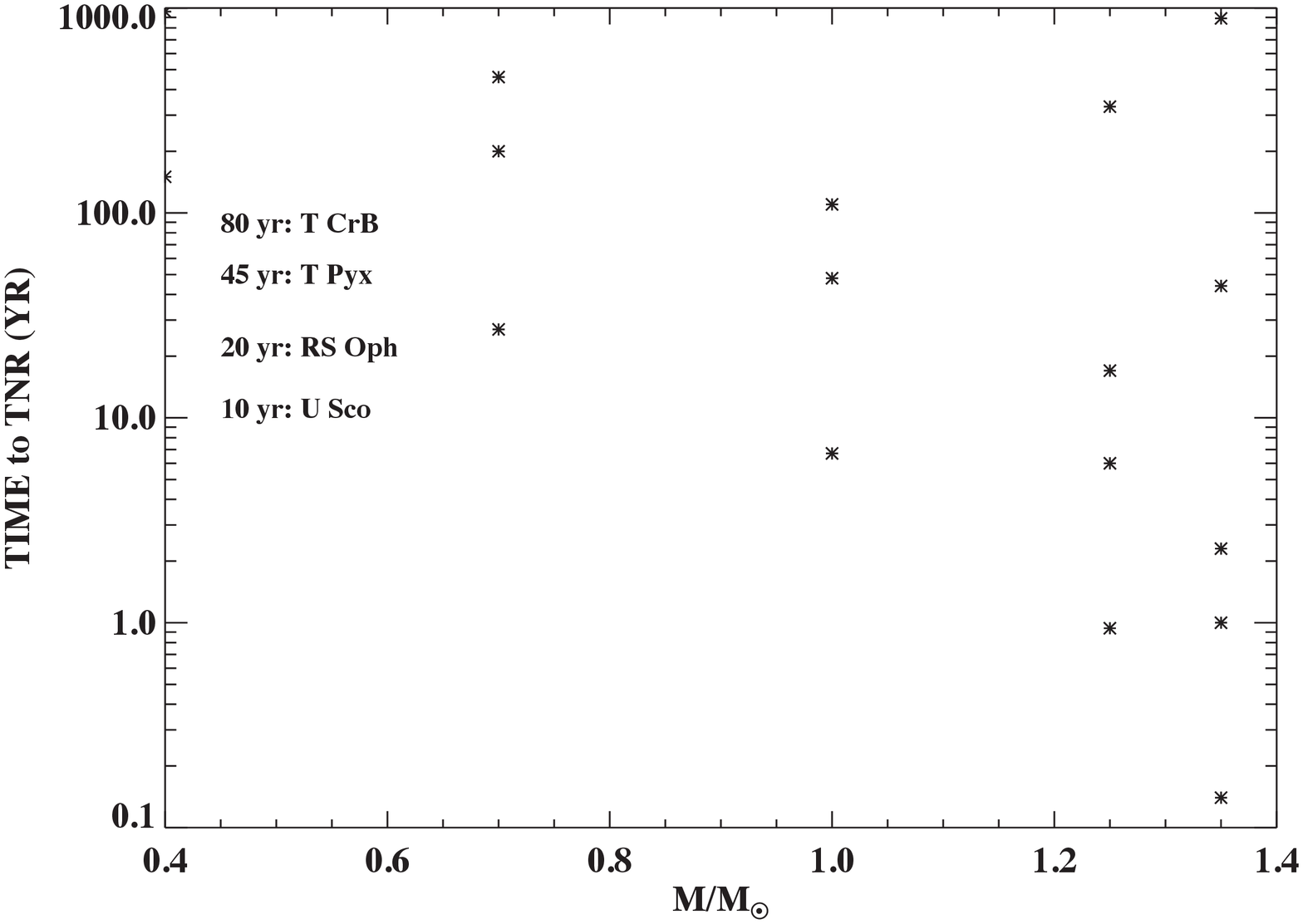,width=100mm}}
\vspace{-5mm}
\captionb{4}
{These are the same data given in Figure 3 but here we concentrate only on the lower right corner and add 
 approximate recurrence times for the best known RNe.  The location of each RN on the left hand side of
 the plot indicates
 its approximate recurrence time.  This plot shows that not only is it possible for
 RNe outbursts to occur on low mass WDs but they can also occur for a broad range of \.M on higher mass WDs.}}
\vspace{-2mm}

\end{figure}

Therefore, we can test the assumptions and results of Fujimoto (1982a,b) by accreting {\it solar material} 
onto the WD, assume no mixing of accreted matter with core matter has occurred, and that this is the ``first'' outburst on 
the WD.  In later studies we will relax some of these assumptions but our fundamental result is that ``steady burning''
does not occur, all evolutionary sequences produce TNRs.  It appears that the semi-analytical study of Fujimoto (1982a,b)
breaks down when compared to fully time-dependent calculations with our modern hydro code.

\begin{figure}[htb!]
\vspace{-5mm}
\vbox{
\centerline{\psfig{figure=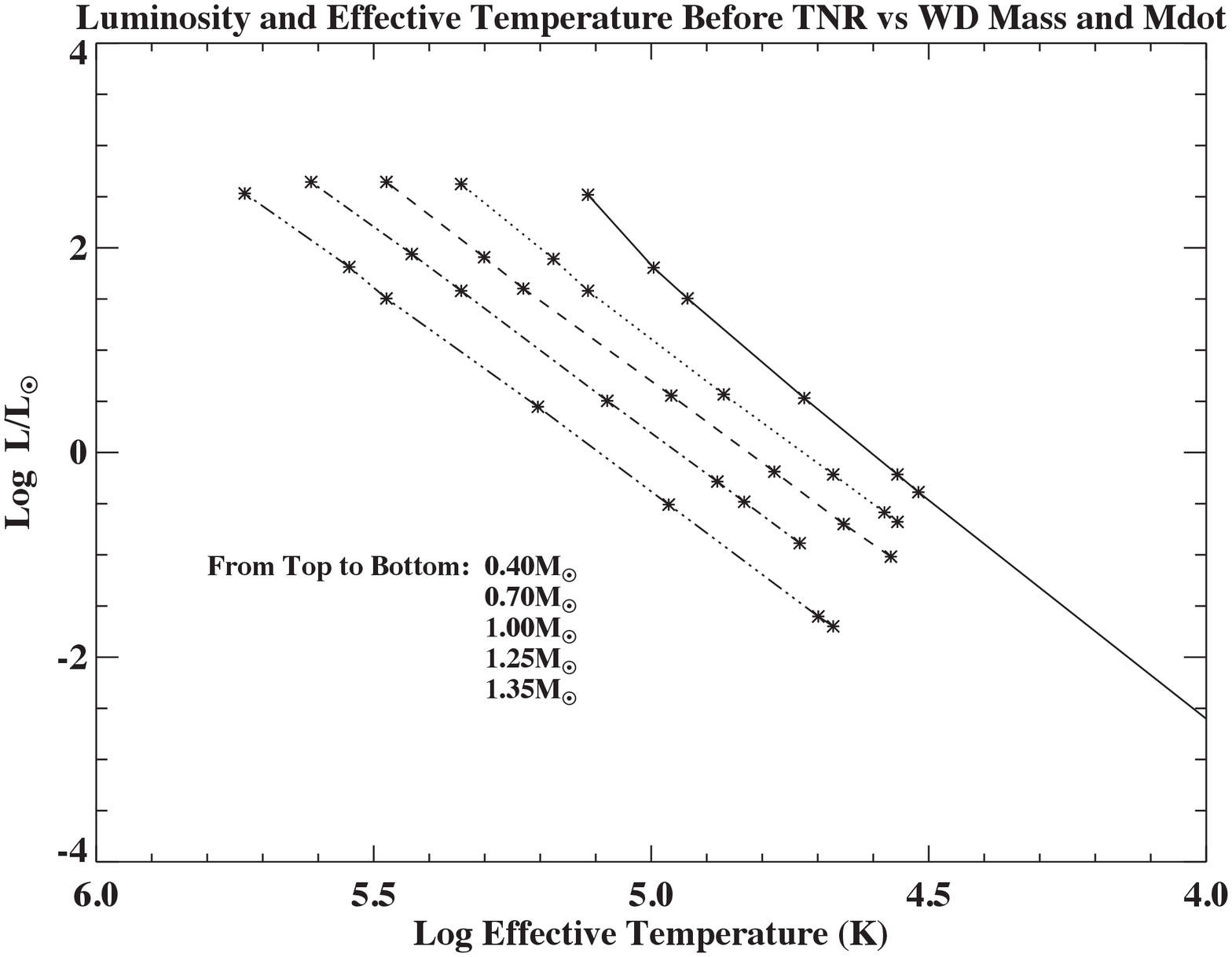,width=100mm}}
\vspace{-5mm}
\captionb{5}
{ We show the luminosity and effective temperature in the HR diagram for our simulations around the time
just before the final thermonuclear runaway.  The mass accretion rate, for all masses, increases from 
low effective temperatures to high effective temperatures.  Most of these sequences, especially the ones occurring on
the lower mass WDs would not be detected by the current low energy detectors on the X-ray satellites.}}
\end{figure}

\begin{figure}[hbt!]
\vspace{-5mm}
\centering
\begin{tabular}{cc}
\epsfig{file=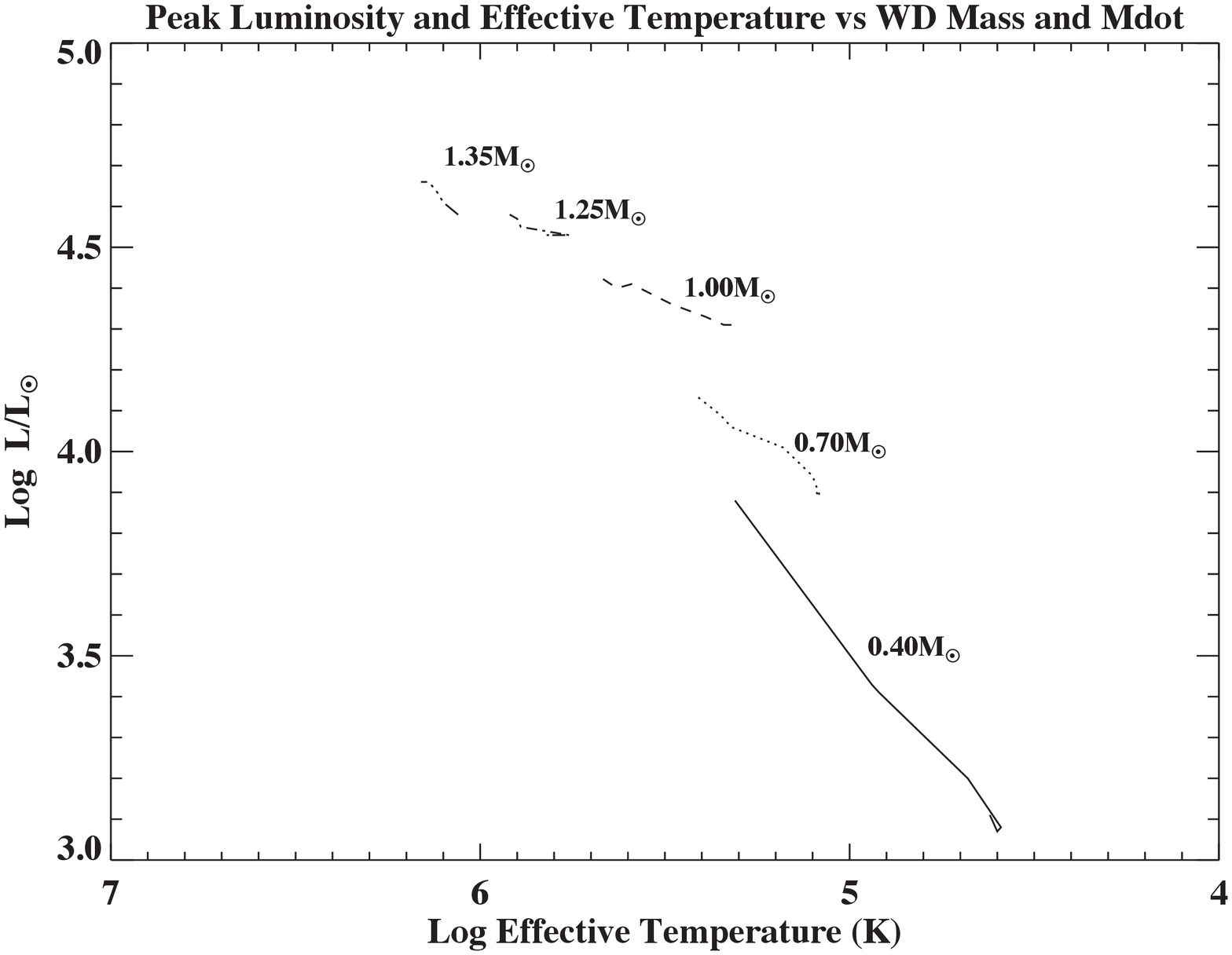,width=100mm} 
\vspace{-5mm}
\end{tabular}
\captionb{6}{This is the same plot as in Figure 5 but for the peak conditions during the thermonuclear runaway. 
Again, the mass accretion rate is increasing from lower effective temperatures to higher effective temperatures.
All but those occurring on the lowest mass WDs would be detected by current X-ray satellites but those on the
highest mass WDs, and thus closest to the Chandrasekhar Limit, would be detected for the shortest amount
of time.}
\end{figure}

\sectionb{4}{RESULTS}

Figure 1 shows the results for all 70 simulations that we have done (each data point represents the two initial
luminosities).  In all cases we obtain a TNR which, for some simulations, ejects some material, and after some
evolutionary time may cause the WD radius to grow to $\sim 10^{12}$cm.  
In no case does ``steady burning'' as described in the literature occur.  These
fully time-dependent calculations show that the sequences exhibit the Schwarzschild \& H\"arm (1965) thin shell instability
which precludes ``steady burning.''   We also find that low mass WDs do not eject any mass while the high mass WDs do eject 
a small fraction of the accreted material (a maximum of $\sim4\%$ for the 1.25M$_\odot$  sequences but ranging down to $\sim0.1\%$ for the 0.7M$_\odot$ sequences).

\begin{figure}[]
\vspace{-10mm}
\centering
\begin{tabular}{cc}
\epsfig{file=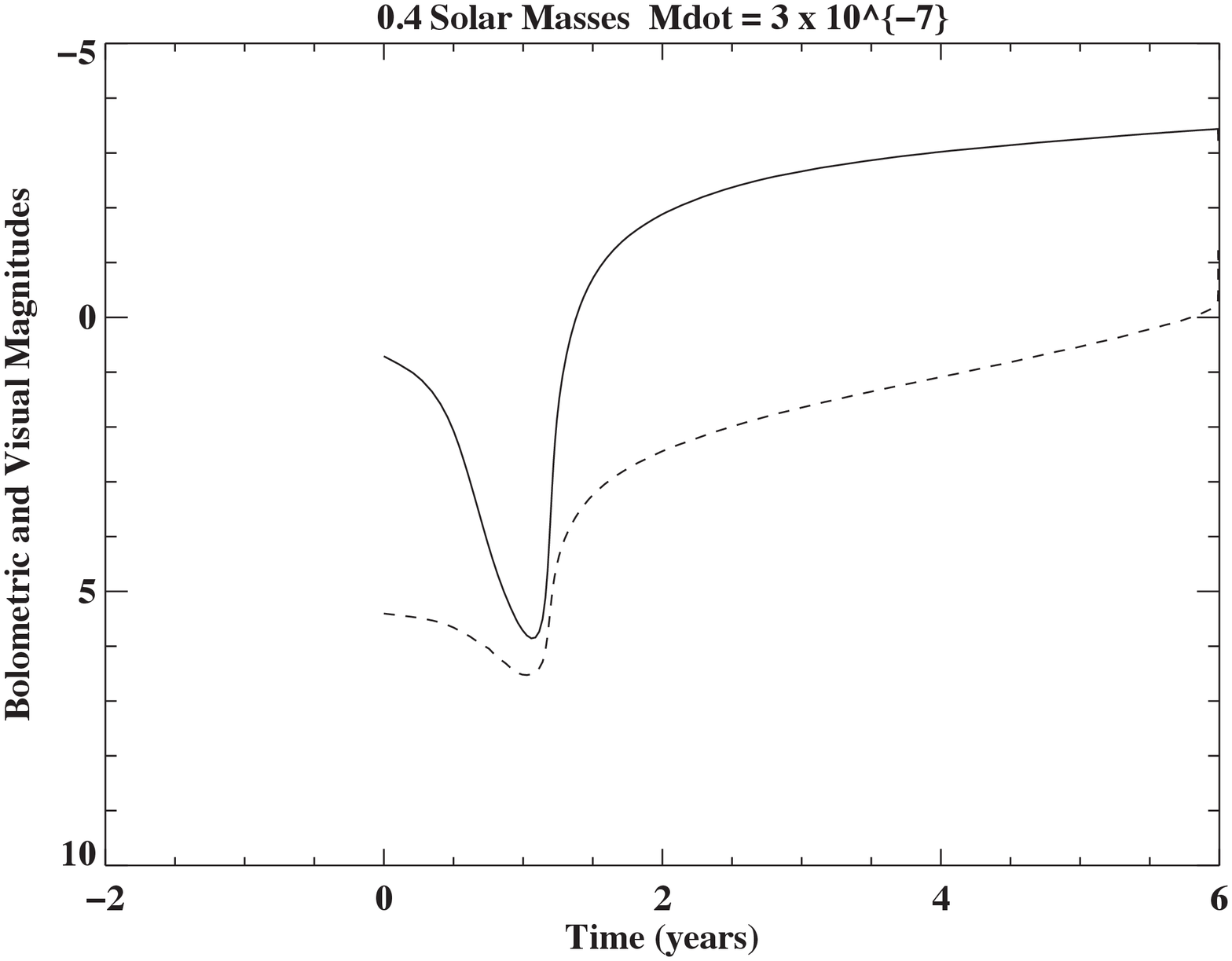,width=100mm}
\end{tabular}
\vspace{-5mm}
\captionb{7}{This figure and the next three figures show the light curves for 4 of our 70 simulations. 
The WD mass and \.M are listed on 
the top of each figures.  They are for four different masses but the same \.M, $3 \times 10^{-7}$M$_\odot$yr$^{-1}$, and show that
choosing an \.M in the middle of the ``steady burning'' regime still results in a thermonuclear runaway.  The dip before the final
rise to maximum is explained in the text.  Note that the simulation on the lowest mass WD, 0.4M$_\odot$, takes years to reach
the peak while that on the highest mass WD, 1.35M$_\odot$ takes less than one day. }
\end{figure}

The crucial difference between these studies and simulations of Classical Novae (CNe) is that
here we assume {\em no} mixing of core matter with accreted matter.  Figure 2, which is the most important result of this paper,  shows the
log of the mass accreted (minus mass lost) as a function of WD mass for each simulation.  It shows that all these WDs are growing in mass as a result of the
accretion of {\it solar material}.   This is, as well known, not the case for CNe which show sufficient core
material in their ejecta that the WD must be losing mass as a result of the outburst.  In CNe, it has long been realized that there
must be mixing of the accreted material with core material and, for most objects, the WD is losing mass as a
result of the nova outburst  (Gehrz et al. 1998).  

Our results now suggests that we need to identify the mixing mechanism,
when it takes place, and how much material from the core is mixed up into the accreted material in order to better understand the
differences between CNe and other systems that contain accreting WDs.  Therefore, we identify our simulations with
those Symbiotic and Recurrent Novae that do not show core material in their ejecta.  It is also likely that for typical Cataclysmic Variables (Dwarf
Novae and related objects which undergo accretion driven outbursts but the WD component must still be accreting material from the secondaries),  which show only a small amount of ejected material, that the WD is growing in mass.   
This may explain the results shown by G\"ansicke in the proceedings  of IAU 281 (see
Zorotovic, M., et al.  2011).   

We also show the log of the accretion time to TNR for all our sequences (Figure 3).  Clearly,
as the WD mass increases, the accretion time decreases for the same \.M. This is well known since
higher mass WDs initiate the TNR with a smaller amount of accreted mass than lower mass WDs.

\begin{figure}[]
\vspace{-10mm}
\centering
\begin{tabular}{cc}
\epsfig{file=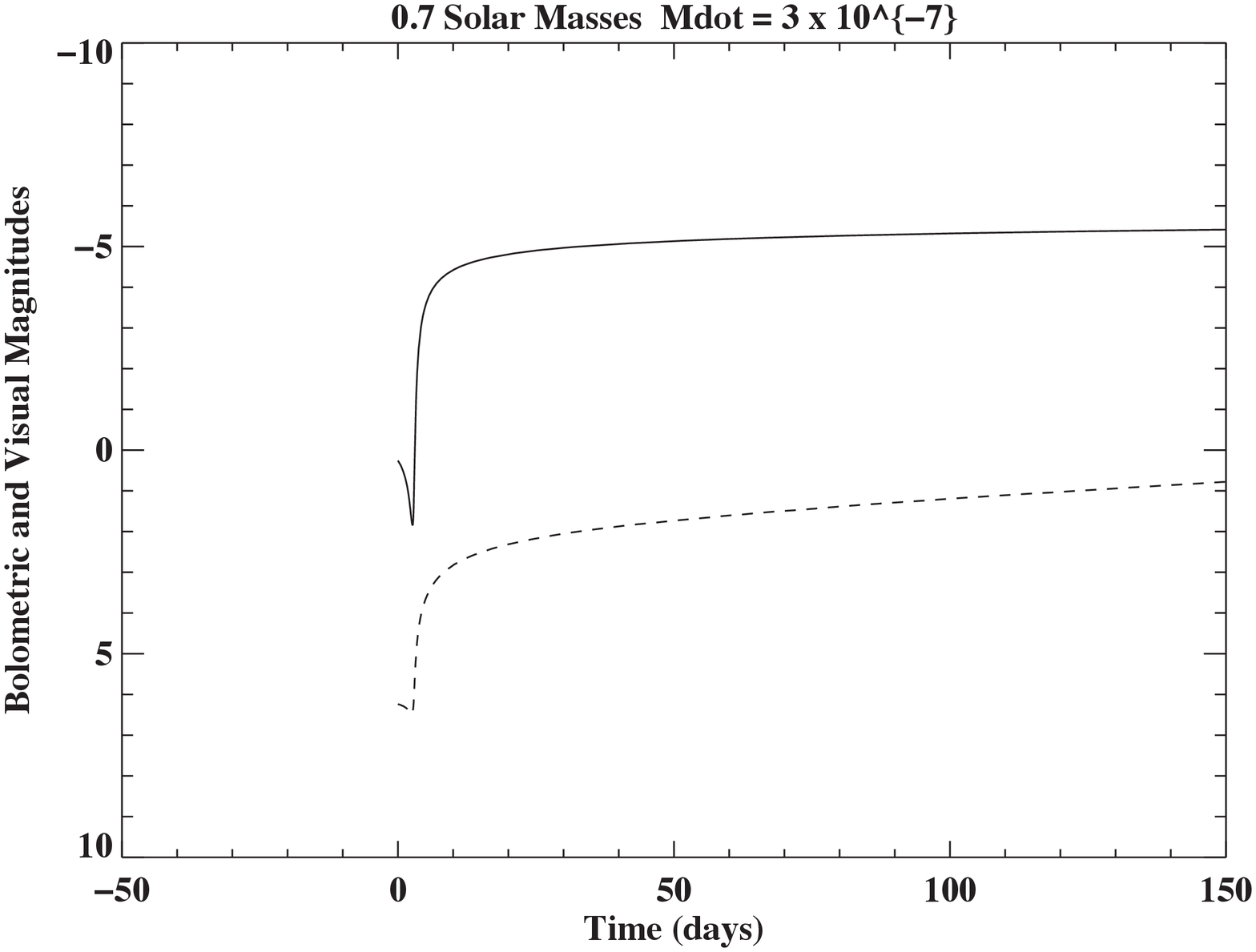,width=100mm}
\end{tabular}
\vspace{-5mm}
\captionb{8}{See caption for Figure 7.}
\end{figure}

In Figure 4, we
magnify the lower right corner of Figure 3 and add approximate recurrence times for the best known RNe.
Although it is often claimed that only the most massive WDs have recurrence times short enough to agree with these
RNe, this plot shows that this is not the case.  It is possible for some of the observed RNe to occur on WDs with masses as
low as 0.7M$_\odot$.  Therefore, determining the mass of the WD in these RNe, based only on short recurrence times, is dangerous.  We also
note that it is possible for a RN outburst to occur on a high mass WD for an extremely broad range of \.M.
A clue to the WD mass is the X-ray emission at maximum since it is likely that X-ray emission observed 
at maximum can only occur on massive WDs. 

Another important result addresses claims that there are insufficient SD systems identified
in various X-ray searches for them to be SN Ia progenitors (c.f., Gilfanov \& Bogd\'an 2010).  We investigated
this question by tabulating the effective temperatures and luminosities of our simulations both during the evolution to
the TNR (Figure 5) and at the peak of the TNR (Figure 6).   To better understand the implications of these plots, we refer
to the evolution of RS Oph in X-rays (Osborne et al. 2011).  Osborne et al. report on the Swift X-ray light curve of 
this Recurrent Symbiotic nova and find that the nova did not become a Super Soft Source until about day 26
to 30 of the outburst.  They interpret this behavior as the consequences of nuclear burning on the surface of
the WD causing its effective temperature to gradually increase until it becomes sufficiently hot for the emission
to be detected by the Swift satellite.  They used an expression that relates the decrease in V magnitude (from peak V) to the
effective temperature (Bath and Harkness 1989), to estimate that, when it was detected by Swift, the temperature of the
nuclear burning WD had reached 500,000K  about 30 days after discovery.  This temperature is much higher than the
black-body temperatures assumed for the Super Soft Sources but agrees with analyses of X-ray
grating spectra obtained at about the same time (Ness et al. 2007).   Given the low energy response of the
ACIS (on Chandra) and Swift XRT, we estimate that a WD evolving to a TNR must exceed an effective
temperature of at least 400,000K before it will be detected by these X-ray satellites.  Figure 5 shows that only the
most massive WDs, accreting at the highest mass accretion rates, could be detected as Super Soft Sources by
the current X-ray satellites.  However, these systems also have the shortest ``duty'' cycles and could be missed
on their evolution to explosion.  So, their non-detection is not surprising.

\begin{figure}[]
\vspace{-10mm}
\centering
\begin{tabular}{cc}
\epsfig{file=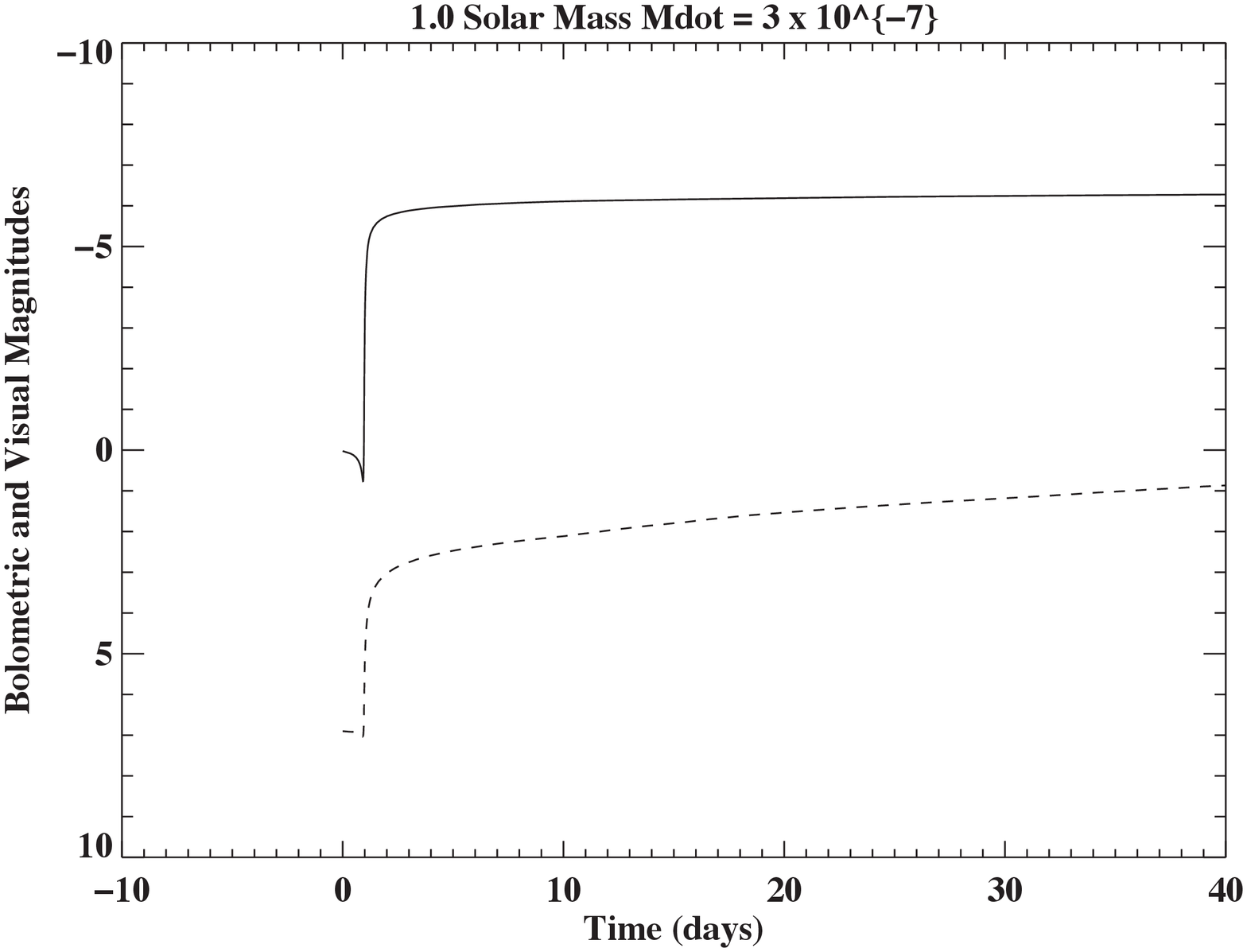,width=100mm}
\end{tabular}
\vspace{-5mm}
\captionb{9}{See caption for Figure 7.}
\end{figure}

Figure 6 shows these systems at their peak in the HR diagram.  The sequences that are the hottest and most
luminous are again those with the highest mass accretion rate at each WD mass.  They are the sequences that
have accreted the least amount of material and, therefore, have ejected the least amount of material.  They will
be ``bright'' in X-rays for the shortest amount of time.  

Finally, in Figures 7 through 10 we show the light curves for four of our simulations.  The WD mass is given on top of each plot
and they are all for the same \.M: $3 \times 10^{-7}$M$_\odot$yr$^{-1}$.  The evolution of the bolometric
magnitude is the solid line in each plot and the V magnitude is shown as the dashed line.   We stop the plot when
the outer radius reaches $\sim 10^{12}$cm and the expanding material becomes optically thin.  
Figure 7 shows the evolution for the 0.4M$_\odot$ sequence and the horizontal axis is in
years.  In the following Figures, we give the horizontal axis in days.  As the mass of the WD increases, the time
scale decreases.  The initial decline before the final rise is caused by the conversion of some of the
internal energy, produced by ongoing nuclear burning near the surface, 
being transformed into the potential energy necessary for the material to climb out of the 
deep gravitational well of the WD.  The most extreme result is for the 0.4M$_\odot$ WD for which
it takes more than one year for the expanding material to recover and begin to become more 
luminous and hotter.  Such an effect, combined with the interaction with the accretion disk 
(and possibly the secondary) might be responsible for the pre-maximum halt seen in some
novae (Hounsell et al. 2010).

\begin{figure}[]
\vspace{-5mm}
\centering
\begin{tabular}{cc}
\epsfig{file=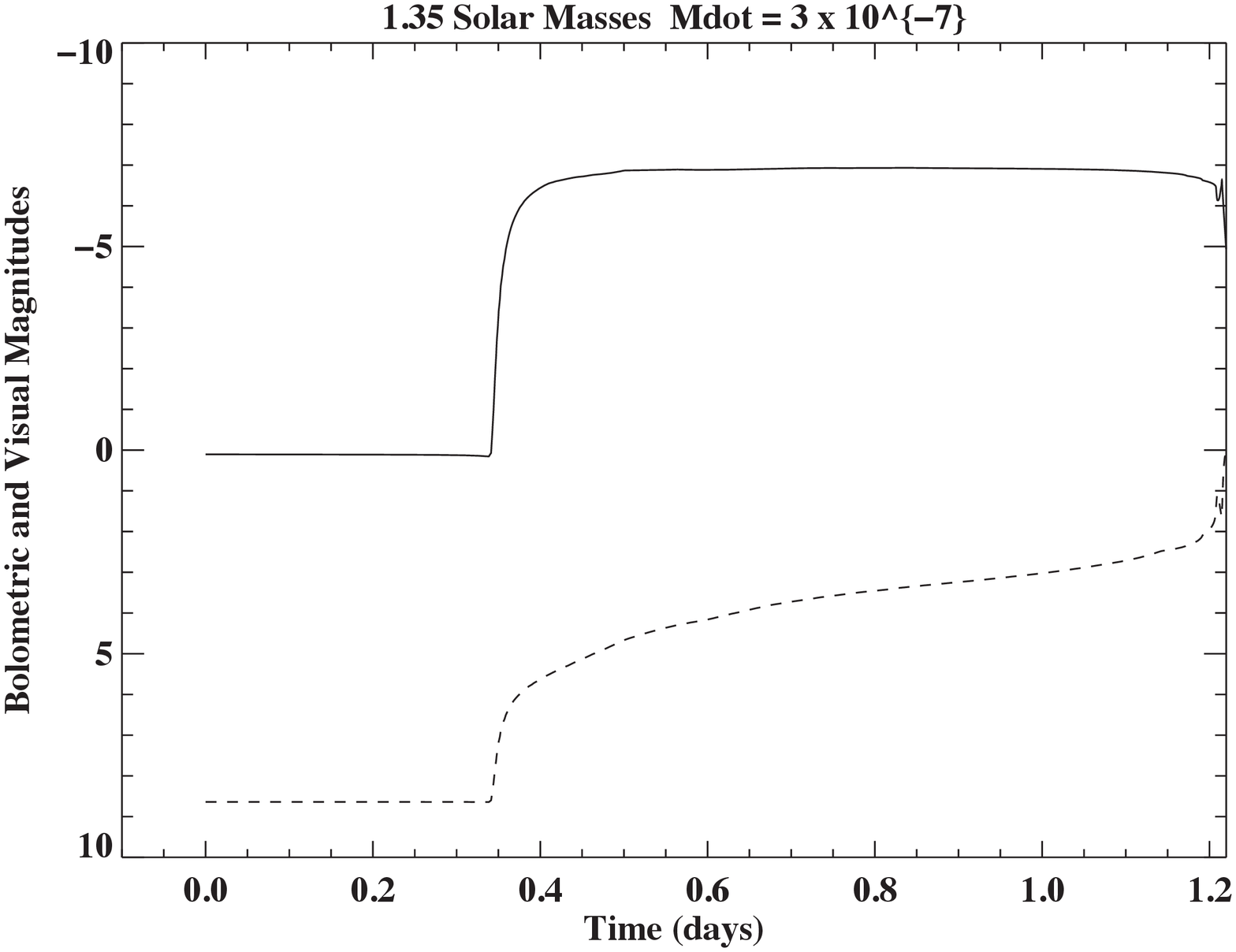,width=100mm}
\end{tabular}
\vspace{-5mm}
\captionb{10}{See caption for Figure 7.}
\end{figure}

\sectionb{5}{CONCLUSIONS}

\begin{itemize}

\item We report that our simulations of accretion of solar material onto WDs always produce a thermonuclear
runaway and ``steady burning'' does not occur.

\item Thermonuclear runaways on WDs more massive than 0.4M$_\odot$ eject only a small fraction of 
the amount accreted to initiate the runaway.  We find the maximum ejected material ($\sim4\%$)
for the 1.25M$_\odot$  sequences and  the amount of ejected gas decreases to $\sim0.1\%$ for the 0.7M$_\odot$ sequences.

\item All WDs are growing in mass as a consequence of the accretion of solar material.

\item The time to runaway is sufficiently short for accretion onto most of the WD masses that we
studied that Recurrent Novae could occur on a much broader range of WD mass than heretofore
believed.

\item During most of the evolution time to the peak of the thermonuclear runaway the surface 
conditions of the WD (effective temperature and luminosity) are too low to be detected by the
currently orbiting low energy X-ray detectors.  The non-discovery of a large number of accreting WDs is not surprising. 

\end{itemize}

\thanks{We gratefully acknowledge partial support from the U. S. National Science Foundation, NASA, and the DOE.}

\vspace{2.0cm}

\References


\refb Arnett, W. D., Meakin, C., and Young, P. A. 2010, {Ap.J.}, 710, 1619

\refb Bath, G., T.  \& Harkness, R. P.  1989, in Classical Novae, ed. M. Bode \&A. Evans,
Wiley Interscience, 61

\refb Branch, D. et al. 1995, PASP, 107, 1019

\refb Fujimoto, M. Y. 1982a, Ap.J., 257, 752

\refb Fujimoto, M.Y. 1982b, Ap. J., 257, 767

\refb Gilfanov, M., \& Bogd\`an, \`A.  2010, NATURE, 463, 924

\refb Hillebrandt, W., \& Niemeyer, J. 2000, ARAA, 38, 191

\refb Hounsell, R., et al. 2010, ApJ, 724, 480

\refb Howell, D. A. 2011, Nature Comm., 2, 350

\refb Iliadis, C., et al. 2010, Nuclear Physics A, 841, 31

\refb Kahabka, P. \& van den Heuvel, E. P. J. 1997, ARAA, 35, 69

\refb Kasen, D., R\"opke, F. K., \& Woosley, S. E. 2009, NATURE, 460, 869

\refb Kasliwal, M. M., et al. 2011, ApJ, 735, 94

\refb Khokhlov,  A. M. 1991, A\&A, 245, 114

\refb Kromer, M. et al. 2010, ApJ, 719, 1067

\refb Leibundgut, B. 2000, A\&A Reviews, 10, 179

\refb Leibundgut, B. 2001, ARAA, 39, 67

\refb Lodders, K., 2003, ApJ, 591, 1220

\refb Mazzali, P. A. et al. 2007, Science, 315, 825

\refb Meakin, C., \& Arnett, W. D. 2007, ApJ, 667, 448

\refb Ness, J.-U., et al. 2007, ApJ, 665, 1334

\refb Nomoto, K. 1982, Ap. J., 253, 798

\refb Nomoto, K., et al. 2000, in COSMIC EXPLOSIONS, AIP Conference Proceedings, Volume 522, 35

\refb Osborne, J. P., et al. 2011, ApJ, 727, 124

\refb Schwarzschild, M. \& H\"arm, R. 1965, ApJ, 142, 855

\refb Starrfield, S. 1989, in Classical Novae, ed. M. Bode \& A. Evans, Wiley Interscience, 39

\refb Starrfield, S., Iliadis, C., \& Hix, W. R. 2008,  in Classical Novae II, ed. M. Bode \& A. Evans,
Cambridge University Press, 77 

\refb  Starrfield, S., Iliadis, C., Hix, W. R., Timmes, F. X., Sparks, W.
M. 2009, ApJ, 692, 1532 

\refb Tornamb\`e, A. \& Piersanti, L. 2005, ASP conference, 342, 169

\refb Van den Heuvel, E. P. J., et al. ,1992, A\&A, 262, 97

\refb Whelan, J., \& Iben, I. 1973, ApJ, 186, 1007

\refb Woosley, S. E. \& Kasen, D. 2011, Ap.J., 734, 38

\refb  Woudt, P. A., et al. 2009, Ap.J, 706, 738

\refb  Zorotovic, M., Schreiber, M.R., G\"ansicke, B.T., 2011 A\&A, 536, A42 


\end{document}